\documentclass[prl,aps,showpacs,epsf,twocolumn]{revtex4}
\usepackage{amssymb}
\usepackage{bm}
\usepackage[dvips]{graphicx}

\def\be{\begin{equation}}
\def\ee{\end{equation}}

\begin{document}

\title{Universal phase diagram of a strongly interacting Fermi gas with unbalanced spin populations.}
\author{F. Chevy}
\affiliation{Laboratoire Kastler Brossel, \'Ecole normale
sup\'erieure, Paris, France }

\date{\today}

\begin{abstract}
We present a theoretical interpretation of a recent experiment
presented in ref. \cite{Zwierlein06} on the density profile of
Fermi gases with unbalanced spin populations. We show that in the
regime of strong interaction, the boundaries of the three phases
observed in \cite{Zwierlein06} can be characterized by two
dimensionless numbers $\eta_\alpha$ and $\eta_\beta$. Using a
combination of a variational treatment and a study of the
experimental results, we infer rather precise bounds for these two
parameters.
\end{abstract}


\maketitle

\section{Introduction}

In fermionic systems, superfluidity arises from the pairing of two
particles with opposite spin states, a scenario first pointed out
by Bardeen Cooper and Schrieffer (BCS) to explain the onset of
superconductivity in metals. For this mechanism to be efficient,
the Fermi surfaces associated with each spin component need be
matched, and soon after the seminal BCS work the question of the
effect of a population imbalance between the two states was
raised. At the time, it was understood that pairing and
superfluidity could sustain a certain amount of mismatch, above
which the system would undergo a quantum phase transition towards
a normal state  \cite{Clogston62}. The original work of Fulde,
Ferrel, Larkin and Ovchinnikov, who proposed the existence of
Cooper pairing at finite momentum, was later generalized to
trapped systems \cite{Combescot05}. Alternative scenarios were
also proposed, including deformed Fermi surfaces
\cite{Sedrakian05}, interior gap superfluidity \cite{Liu03}, phase
separation between a normal and a superfluid state through a first
order phase transition \cite{Bedaque03}, BCS quasi-particle
interactions \cite{Ho06} or onset of p-wave pairing
\cite{Bulgac06}. When the strength of the interactions is varied,
a complicated phase diagram mixing several of these scenarios is
expected \cite{Pao05}.

However, due to the absence of experimental evidence, these
scenarios could never be tested experimentally until the subject was
revived by the possibility of reaching superfluidity in ultra-cold
fermionic gaseous systems \cite{Jochim03,Bourdel04}. Contrarily to
usual condensed matter systems, spin relaxation is very weak in cold
atoms, and this allows one to keep spin polarized samples for long
times. This unique possibility led to the first experimental studies
of imbalanced Fermi gases at MIT and Rice University
\cite{Zwierlein05,Partridge05,Zwierlein06}. These results triggered
a host of theoretical work aiming at explaining the various results
observed by the two groups \cite{Pieri05,Chevy06}.

One remarquable feature of ref. \cite{Zwierlein06} is the
observation of three different phases in the cloud. At the center,
the authors observe a superfluid core, where the densities of the
two spin states are equal, then an intermediate normal shell where
the two states coexist and finally an outer rim of the majority
component. In the present paper, we show that, though performed in
a trap, the observations of MIT can offer valuable information on
the phase diagram of a strongly interacting Fermi gas with
unbalanced populations. In a first part we will present a brief
overview of the simplest free space scenario for transition from a
paired superfluid to a pure normal state, through a mixed phase.
Focusing on the disappearance of the minority component, we will
present a variational study of the problem of a single minority
particle embedded in the Fermi sea of majority atoms. Finally, we
will show that the comparison with experiments allows for a rather
precise determination of the transition thresholds. One of the key
point is that, contrarily to previous works, we rely on universal
thermodynamics \cite{Ho03} as well as ``exact" experimental or
Monte Carlo results, without the need of the mean field BCS ansatz
often used in other publications, an approach similar to that of
\cite{Bulgac}.

\section{Homogeneous system}

Before addressing the case of trapped fermions, let us first
discuss their free space phase diagram. We consider zero
temperature fermions of mass $m$ with two internal states labelled
1 and 2. Within a quantization volume $V$ and in the limit of
short range interactions, we can write the hamiltonian of the
system as

\begin{equation}
\widehat H=\sum_{\bm k,\sigma}\epsilon_{\bm k}\widehat a_{\bm
k,\sigma}^\dagger \widehat a_{\bm k,\sigma}+\frac{g_{\rm
b}}{V}\sum_{\bm k,\bm k',\bm q}\widehat a^\dagger_{\bm k+\bm q,1}
\widehat a_{\bm k'-\bm q,2}^\dagger \widehat a_{\bm k',2} \widehat
a_{\bm k,1}.
\end{equation}

 Here, $\epsilon_{\bm k}=\hbar^2 k^2/2m$, $\widehat
a_{\bm k,\sigma}$ is the annihilation operator of a species
$\sigma$ particle with momentum $\bm k$, and $g_{\rm b}$ is the
bare coupling constant characterizing inter-particle interactions.
It is related to the s-wave scattering length of the system $a$ by
the Lippmann-Schwinger equation

\begin{equation}
\frac{1}{g_{\rm b}}=\frac{m}{4\pi\hbar^2 a}-\frac{1}{V}\sum_{\bm
k}\frac{1}{2\epsilon_{\bm k}}. \label{Eqn5}
\end{equation}

\noindent We note that only interactions between particles of
opposite spins are taken into account, due to the Pauli principle
which forbids s-wave scattering of atoms with identical spin. In
this paper, we assume we are working at the unitary limit where
$|a|=\infty$ and, using the grand canonical ensemble, we wish to
find the ground state of the grand-potential $\widehat \Xi=\widehat
H-\mu_1 \widehat N_1-\mu_2\widehat N_2$. Here $\widehat N_i$ is the
particle number operator for species $i$ and $\mu_i$ is the
associated chemical potential (we take species 1 as majority hence
$\mu_1>\mu_2$). The grand potential can be expressed as the function
of the volume $V$ and pressure $P$ of the ensemble according to
$\langle\widehat \Xi\rangle=-PV$. In other words, searching the
ground state of the system is equivalent to searching the phase with
the highest pressure $P$.

To start our analysis, we note first that two exact eigenstates of
$\widehat\Xi$ can be found quite easily. First, when the gas is
fully polarized, we recover the case of an ideal Fermi gas for
which we know that

$$P_{\rm
N}=\frac{1}{15\pi^2}\left(\frac{2m}{\hbar^2}\right)^{3/2}\mu_1^{5/2}.$$

Second, let us now consider the exact ground state $|{\rm
SF}\rangle_\mu$ of the balanced grand potential $\widehat
\Xi'=\widehat H-\mu (\widehat N_1+\widehat N_2)$, describing a
superfluid with chemical potential $\mu$. This potential commutes
with the number operators, hence the ground state can be searched
as an eigenstate for both $\widehat N_1$ and $\widehat N_2$, with
$\widehat N_1|{\rm SF}\rangle_\mu=\widehat N_2|{\rm
SF}\rangle_\mu$. We check readily that $|{\rm SF}\rangle_{\bar
\mu}$, with $\bar\mu=(\mu_1+\mu_2)/2$, is still an eigenstate of
the unbalanced $\widehat \Xi$, by noting that
$\widehat\Xi=\widehat\Xi'+(\mu_1-\mu_2)(\widehat N_1-\widehat
N_2)/2$. At unitarity the pressure of a balanced Fermi gas reads
$P_{\rm S}=2\left(2m/\xi\hbar^2\right)^{3/2}\mu^{5/2}/15\pi^2$,
where $\xi\sim 0.42$ is a universal parameter  whose determination
has attracted interest of both theoreticians
\cite{Carlson03,Perali04,Carlson05,Astrakharchik04} and
experimentalists \cite{Bourdel04,OHara02}. In the case of
mismatched chemical potentials, the pressure of this fully paired
superfluid state is therefore

$$P_S=\frac{1}{15\pi^2}\left(\frac{m}{\xi\hbar^2}\right)^{3/2}(\mu_1+\mu_2)^{5/2}.$$

\begin{figure}
\includegraphics[width=\columnwidth]{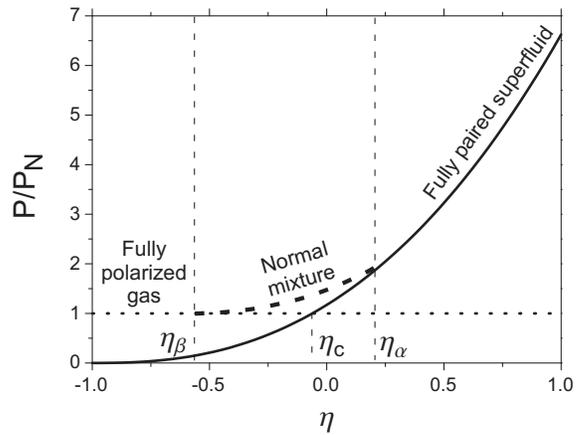}
\caption{Comparison of the pressure of the various phases,
normalized by the pressure $P_N$ of the fully polarized ideal
Fermi gas, as a function of the chemical potential mismatch
$\eta=\mu_2/\mu_1$. Dotted line: fully polarized phase. Full line:
fully paired superfluid phase. The fully paired and fully
polarized states meet for $\eta_c\sim -0.099$. Dashed line: sketch
of the intermediate normal phase. $\eta_\alpha$ and $\eta_\beta$
designate the universal chemical potential thresholds for this
phase.} \label{Fig1}
\end{figure}

The evolution of $P_N$ and $P_S$ is presented in Fig. \ref{Fig1} as
a function of $\eta=\mu_2/\mu_1$. We see that they cross for
$\eta_{\rm c}=(2\xi)^{3/5}-1\sim -0.10$, marking the instability of
the superfluid against large population imbalances \cite{Chevy06}.
However, since we only compare the energy of the fully paired state
to the one of the fully polarized ideal gas, the real breakdown of
superfluidity could very well happen for some $\eta$ larger than
$\eta_c$. We know this is actually the case, since in ref.
\cite{Zwierlein06}, the authors observed an intermediate normal
phase, containing atoms of both species. From universality at
unitarity, the phase transition from the fully paired to the
intermediate phase, and then from the intermediate to the fully
polarized normal phase are given by conditions
$\mu_2/\mu_1=\eta_\alpha$, and $\mu_2/\mu_1=\eta_\beta$, where
$\eta_\alpha$ and $\eta_\beta$ are two universal parameters we would
like to determine as precisely as possible.

Noting that the transition from the fully paired state to the
intermediate one must happen before the transition to the fully
polarized phase, we see graphically that we have necessarily
$\eta_\alpha>\eta_c$, and, similarly, $\eta_\beta<\eta_c$. The
upper bound on $\eta_\beta$ can be further improved by noting that
at the threshold between the normal mixture and the fully
polarized ideal gas, there are only a few atoms of the minority
species. In principle, the value of $\eta_\beta$ should then be
found by studying the N+1 body problem of a Fermi sea of N
particles 1, in presence of a single minority atom. To address
this problem, we use here a variational method inspired from first
order perturbation theory, where we expect the ground state of the
system to take the form

$$|\psi\rangle=\phi_0|{\rm FS}\rangle+\sum_{\bm k,\bm q}\phi_{\bm k,\bm
q}|\bm k,\bm q\rangle,$$

\noindent where $|{\rm SF}\rangle$ is a non interacting majority
Fermi sea plus a minority atom with 0 momentum, and $|\bm k,\bm
q\rangle$ is the perturbed Fermi sea with a majority atom with
momentum $\bm q$ (with $q$ lower than $k_F$) excited to momentum
$\bm k$ (with $k>k_F$). To satisfy momentum conservation, the
minority atom acquires a momentum $\bm q-\bm k$. The energy of
this state with respect to the non interacting ground state is
$\langle \widehat \Delta H\rangle=\langle\widehat
H_0\rangle+\langle\widehat V\rangle$, with

$$\langle\widehat H_0\rangle=\sum_{\bm k,\bm q}|\phi_{\bm k,\bm
q}|^2(\epsilon_{\bm k}+\epsilon_{\bm q-\bm k}-\epsilon_{\bm q}),$$

\noindent and

\begin{eqnarray*}
\langle\widehat V\rangle&=&\frac{g_b}{V}\left(\sum_{\bm
q}|\phi_0|^2+\sum_{\bm k,\bm k',\bm q}\phi_{\bm k',\bm q}\phi^*_{\bm
k,\bm q}+\sum_{\bm k,\bm q,\bm q'}\phi_{\bm k,\bm q}\phi^*_{\bm
k,\bm q'}\right.\\
&&\left. +\sum_{\bm q,\bm k}(\phi^*_0\phi_{\bm k,\bm
q}+\phi_0\phi^*_{\bm k,\bm q})\right),
\end{eqnarray*}

\noindent where the sums on $q$ and $k$ are implicitly limited to
$q<k_F$ and $k>k_F$. As we will check later (see below, eqn.
(\ref{Eqn6})), $\phi_{\bm k,\bm q}\sim 1/k^2$ for large momenta,
in order to satisfy the short range behavior $1/r$ of the pair
wave function in real space. This means that most of the sums on
$\bm k$ diverge for $k\rightarrow\infty$. This singular behavior
is regularized by the renormalization of the coupling constant
using the Lippman-Schwinger formula, thus yielding a vanishing
$g_B$. As a consequence, since the third sum in $\langle\widehat
V\rangle$ is convergent, it gives a zero contribution to the final
energy when multiplied by $g_B$, and can therefore be omitted in
the rest of the calculation.

The minimization of $\langle \widehat H\rangle$ with respect to
$\phi_0$ and $\phi_{\bm k,\bm q}$ is straightforward and yields
the following set of equations

\begin{eqnarray*}
(\epsilon_{\bm k}+\epsilon_{\bm q-\bm k}-\epsilon_{\bm
q})\phi_{\bm k,\bm q}+\frac{g_b}{V}\sum_{\bm k'}\phi_{\bm k',\bm
q}+\frac{g_b}{V}\phi_0&=&E\phi_{\bm k,\bm q},\label{Eqn1}\\
\frac{g_b}{V}\sum_{\bm q}\phi_0+\frac{g_b}{V}\sum_{\bm q,\bm
k}\phi_{\bm k,\bm q}&=&E\phi_0\label{Eqn2}
\end{eqnarray*}

\noindent where $E$ is the Lagrange multiplier associated to the
normalization of $|\psi\rangle$, and can also be identified with
the trial energy. These equations can be solved self consistently
by introducing an auxiliary function $\chi(\bm q)=\phi_0+\sum_{\bm
k}\phi_{\bm k,\bm q}$ and we obtain

\be \phi_{\bm k,\bm q}=\frac{g_B\chi (\bm q)/V}{E-(\epsilon_{\bm
k}+\epsilon_{\bm q-\bm k}-\epsilon_{\bm q})}, \label{Eqn6}\ee

After a straightforward calculation, this yields

$$
E=\sum_{q<k_F}\frac{1}{\sum_{k>k_F}\left(\frac{1}{\epsilon_{\bm
k}+\epsilon_{\bm q-\bm k}-\epsilon_{\bm
q}-E}-\frac{1}{2\epsilon_{\bm
k}}\right)-\sum_{k<k_F}\frac{1}{2\epsilon_{\bm k}}}. \label{Eqn4}
$$

\noindent where we got rid of the bare coupling constant $g_B$ by
using the Lippman-Schwinger equation (\ref{Eqn5}). This equation
can be solved numerically and yields $E=-0.3 \hbar^2 k_F^2/m$,
{\em i.e.} $\eta_\beta<-0.6$ \cite{Lobo06}. Note that the same
analytical result was obtained independently in
\cite{Combescot06}.

\section{Trapped system and comparison with experiments}

In the rest of the paper, we would like to show how experimental
data from ref. \cite{Zwierlein06} permits to improve the
determination of the parameters $\eta_{\alpha,\beta}$ . In this
pursuit, we use the Local Density Approximation (LDA) to calculate
the density profile of the cloud in a harmonic trap, which for
simplicity we assume is isotropic. In this case, the chemical
potentials $\mu_{1,2}$ of each species depends on position according
to the law $\mu_{1,2}(\bm r)=\mu_{1,2}^0-m\omega^2 r^2/2$, where
$\omega$ is the trap frequency.

\begin{figure}
\includegraphics[width=\columnwidth]{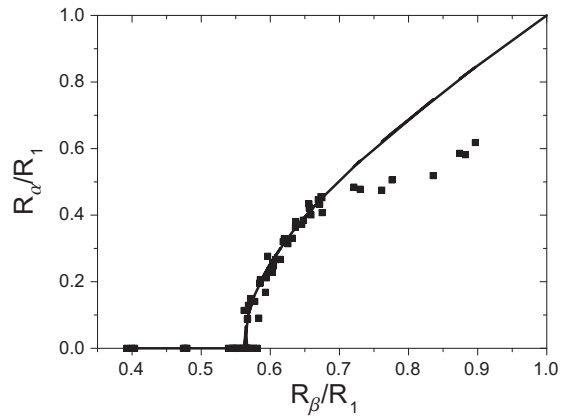}
\caption{Comparison between experimental data of ref.
\cite{Zwierlein06} (squares) and eq. (\ref{Eqn3}) (full line).
$R_\alpha$, $R_\beta$ and $R_1$ are respectively the radii of the
superfluid, minority and majority components. The condensate
vanishes for $R_\alpha=0$ at $q=R_\beta^2/R_1^2\sim 0.32$. }
\label{Fig2}
\end{figure}

Using this assumption, the transition between the various phases
will happen at radii $R_\alpha$ and $R_\beta$ given by
$\mu_2(R_{\alpha,\beta})/\mu_1(R_{\alpha,\beta})=\eta_{\alpha,\beta}$.
When these two equations are associated with the condition giving
the radius $R_1$ of the majority component, $\mu_1(R_1)=0$, we can
eliminate both $\mu_1^0$ and $\mu_2^0$ from the equations,
yielding the following close formula relating the radii
$R_\alpha$, $R_\beta$ and $R_1$

\begin{equation}
\frac{R_\alpha}{R_1}=\sqrt{\frac{(R_\beta/R_1)^2-q}{1-q}}.
\label{Eqn3}
\end{equation}

\noindent Here, $q=(\eta_\alpha-\eta_\beta)/(1-\eta_\beta)$
corresponds to the value of the $R_\beta^2/R_1^2$ at which
$R_\alpha$ vanishes, {\em i.e.} at which the superfluid fraction
disappears. In Fig. \ref{Fig2}, we compare the prediction of eq.
\ref{Eqn3} with the experimental finding of ref.
\cite{Zwierlein06}, taking $q=0.32$ to match the superfluidity
thresholds. We see that close to the threshold, the agreement
between the two graphs is quite good. However, they depart from
each other for $R_\beta/R_1\gtrsim 0.7$, corresponding to low
population imbalance. One explanation for this discrepancy might
involve finite temperature effects. Indeed, it was already noted
in Fig. 4 of ref. \cite{Zwierlein05} that, although the superfluid
fraction was very sensitive to temperature at small imbalances,
the value of the critical population imbalance was more robust.

 The superfluid phase disappears when $R_\alpha$ vanishes.
From eq. \ref{Eqn3}, we see this  happens for a ratio
$R^2_\beta/R^2_1=q$. As seen in Fig. \ref{Fig2}, $q$ can be
extracted from the experimental data of ref. \cite{Zwierlein06},
which yield $q\sim 0.32$ and therefore constrain the possible
values of $\eta_\alpha$ and $\eta_\beta$. Indeed, using this
determination of $q$, as well as the rough upper and lower values
for $\eta_{\alpha,\beta}$, one obtains

\begin{eqnarray}
-0.10<\eta_\alpha<-0.088\\
-0.62<\eta_\beta<-0.60
\end{eqnarray}

These bounds can be compared to the values deduced by BCS theory,
predicting $\eta_\alpha\sim 0.1$ and $\eta_\beta=0$. Our
calculation excludes these values and explains why the width of
the mixed normal state predicted by BCS theory is much narrower
than observed in experiments.

\section{Conclusion}

In conclusion, we have presented an analysis of the experimental
data of ref. \cite{Zwierlein06} providing stringent bounds on the
values of the thresholds for quantum phase transitions in uniform
unbalanced fermi gases. Since they were obtained using minimal
assumptions (mainly zero temperature and LDA), these bounds are
fairly robust. In particular, they do not depend precisely on the
superfluid nature of the intermediate phase. Our results suggest
interesting follow-ups. First, the full understanding of the
system, and in particular of the density profile of the cloud,
requires the knowledge of the state equation of the intermediate
phase, whose exact nature then needs to be clarified. Second, the
comparison with the data of ref. \cite{Partridge05} suggests an
intriguing issue. Indeed, although the superfluidity threshold was
not directly measured in this paper, the parameter $q$ can be
inferred from the critical imbalance 0.7 measured by MIT. Rice's
experimental data yield at this value $q\sim 0.16$. Not only is
this value very far from the one obtained here from the analysis
of MIT's experiments, but it also contradicts the theoretical
bounds $\eta_\alpha>-0.10$ and $\eta_\beta<-0.60$ which imply
$q>0.31$. As suggested in ref. \cite{Silva06b}, this discrepancy
may arise from surface tension effects provoked by the strong
anisotropy of Rice's trap. Another interpretation might be the
onset  of the intermediate phase due to finite temperature
effects, as suggested by some mean-field scenarios. Finally, the
$\eta_{\alpha,\beta}$ parameters can be evaluated experimentally
using the value of $q$ associated with the measurement of the
density discontinuity $\Delta n_{1,2}$ at $r=R_\alpha$,
 given by $\Delta n_1/\Delta n_2=-\eta_\alpha$ \cite{Footnote}. The
preliminary data presented in fig. 2.b of ref. \cite{Zwirlein06c}
suggest that the discontinuity $\Delta n_1$ is very weak, hence
indicating a small value of $\eta_\alpha$, in agreement with the
bounds obtained here.

\acknowledgments The author wishes to thank M. Zwierlein, M.
McNeil-Forbes, A. Bulgac, C. Salomon, F. Werner, L. Tarruell, J.
McKeever and the ENS cold atom group for fruitful discussions. This
work is partially supported by CNRS, Coll\`ege de France, ACI
nanoscience and R\'egion Ile de France (IFRAF). Laboratoire Kastler
Brossel is Unit\'e de recherche de l'\'Ecole normale sup\'erieure et
de l'Universit\'e Pierre et Marie Curie, associ\'ee au CNRS.


\begin{thebibliography}{30}



\bibitem{Clogston62} B.S. Chandrasekhar, Appl. Phys. Lett. {\bf 1 },7
(1962), A. M. Clogston, Phys. Rev. Lett. {\bf 9}, 266 (1962); {G.
Sarma}, {Journal of Physics and Chemistry of Solids} {\bf 24},
1029 (1963); {P. Fulde, R. A. Ferrell}, {Phys. Rev.} {\bf 135},
A550 (1964); {J. Larkin, Y. N. Ovchinnikov}, {Sov. Phys. JETP}
{\bf 20}, 762 (1965).

\bibitem{Combescot05} {R. Combescot}, {Europhys. Lett.} {\bf 55} (2),150
(2001); C. Mora and R. Combescot, Phys. Rev. B.  {\bf 71}, 214504.
(2005); {P. Castorina, M. Grasso, M. Oertel, M. Urban, and D.
Zappal\`a} {Phys. Rev. A} {\bf 72}, 025601 (2005); T. Mizushima,
K. Machida, and M. Ichioka, Phys. Rev. Lett. {\bf 94}, 060404
(2005); T. Mizushima, K. Machida, and M. Ichioka, Phys. Rev. Lett.
{\bf 95}, 117003 (2005); K. Machida, T. Mizushima, and M. Ichioka,
Phys. Rev. Lett. {\bf 97}, 120407 (2006) .

\bibitem{Sedrakian05}{A. Sedrakian, J. Mur-Petit, A. Polls, and H. M\"uther}, {Phys. Rev. A} {\bf 72}, 013613 (2005).

\bibitem{Liu03} {W.V. Liu and F. Wilczek}, {Phys. Rev. Lett.} {\bf
90}, 047002 (2003).

\bibitem{Bedaque03} {P.F. Bedaque, H. Caldas, and G. Rupak}, {Phys.
Rev. Lett.} {\bf 91} 247002 (2003); {H. Caldas}, {Phys. Rev. A}
{\bf 69}, 063602 (2004); {T.D. Cohen}, {Phys. Rev. Lett.} {\bf
95}, 120403 (2005).

\bibitem{Carlson05} {J. Carlson and S. Reddy}, {Phys. Rev.
Lett.} {\bf 95}, 060401 (2005);

\bibitem{Ho06} T.-L. Ho and H. Zai, cond-mat/0602568.

\bibitem{Bulgac06} A. Bulgac, M. McNeil Forbes, and A. Schwenk, Phys. Rev. Lett. {\bf 97}, 020402 (2006).

\bibitem{Pao05} C.H. Pao, Shin-Tza Wu and S.-K. Yip,  Phys. Rev. B {\bf 73}, 132506 (2006); {D.T Son and M.A. Stephanov}, Phys. Rev. A {\bf 74}, 013614 (2006); {D.E. Sheehy and L. Radzihovsky}, Phys. Rev. Lett. {\bf 96},
060401 (2006).


\bibitem{Jochim03}S. Jochim, M. Bartenstein, A. Altmeyer, G. Hendl, S. Riedl, C. Chin, J. H. Denschlag, and R. Grimm,
{Science} {\bf 302}, 2101 (2003); M. W. Zwierlein, C. A. Stan, C.
H. Schunck, S. M. F. Raupach, S. Gupta, Z. Hadzibabic, and W.
Ketterle, {Phys. Rev. Lett.} {\bf 91}, 250401 (2003);
{M.\,Greiner, C.\,A.\,Regal, and D.\,S\,Jin}, {Nature} {\bf 426},
537 (2003); {J. Kinast, S. L. Hemmer, M. E. Gehm, A. Turlapov, and
J. E. Thomas}, {Phys. Rev. Lett.} {\bf 92}, 150402 (2004); {G. B.
Partridge, K. E. Strecker, R. I. Kamar, M. W. Jack, and R. G.
Hulet}, {Phys. Rev. Lett.} {\bf 95}, 020404 (2005).

\bibitem{Bourdel04} T. Bourdel, L. Khaykovich, J. Cubizolles, J. Zhang, F. Chevy, M. Teichmann, L. Tarruell,
S. J. J. M. F. Kokkelmans, and C. Salomon, {Phys. Rev. Lett.} {\bf 93}, 050401 (2004).


\bibitem{Zwierlein05} {M. W. Zwierlein, A. Schirotzek, C. H. Schunck, and W. Ketterle}, Science {\bf 311}, 492 (2006).

\bibitem{Partridge05} {G. B. Partridge, W. Li, R. I. Kamar, Y. Liao, and R. G. Hulet}, Science {\bf 311}, 503 (2006).

\bibitem{Zwierlein06} M. W. Zwierlein, C. H. Schunck, A. Schirotzek, W. Ketterle, Nature {\bf 442}, 54 (2006).


\bibitem{Pieri05} {P. Pieri, G.C. Strinati}, Phys. Rev. Lett. {\bf
96}, 150404 (2006); {W. Yi and L.-M. Duan}, Phys. Rev. A {\bf 73},
031604(R) (2006); T. N. De Silva, and E. J. Mueller, Phys. Rev. A
{\bf 73}, 051602 (R) (2006); M. Haque, and H.T.C. Stoof, Phys.
Rev. A {\bf 74}, 011602 (2006); H. Hu and X.-J. Liu, Phys. Rev. A
{\bf 73} 051603(R) (2006).

\bibitem{Chevy06} F. Chevy,
Phys. Rev. Lett. {\bf 96}, 130401 (2006).

\bibitem{Ho03} T.L. Ho
Phys. Rev. Lett. {\bf 92}, 090402 (2004).

\bibitem{Bulgac} A. Bulgac and M. McNeil Forbes, e-print
cond-mat/0606043.


\bibitem{Carlson03} {J. Carlson, S.-Y. Chang, V. R. Pandharipande, and K. E. Schmidt
}, {Phys. Rev. Lett.} {\bf 91}, 050401 (2003).



\bibitem{Astrakharchik04} {G. E. Astrakharchik, J. Boronat, J. Casulleras, S.
Giorgini}, {Phys. Rev. Lett.} {\bf 93}, 200404 (2004).

\bibitem{Perali04} {A. Perali, P. Pieri, G. C. Strinati}, {Phys. Rev. Lett.} {\bf 93},
100404 (2004).


\bibitem{OHara02} K. M. O'Hara, S. L. Hemmer, M. E. Gehm, S. R.
Granade, and J. E. Thomas, {Science} {\bf 298}, 2179 (2002);  M.
Bartenstein, A. Altmeyer, S. Riedl, S. Jochim, C. Chin, J. Hecker
Denschlag, and R. Grimm, {Phys. Rev. Lett.} {\bf 92}, 120401
(2004); J. Kinast, A. Turlapov, J. E. Thomas, Q. Chen, J. Stajic,
and K. Levin, Science {\bf 307}, 1296 (2005).

\bibitem{Silva06b} T.N. De Silva and E.J. Mueller, Phys. Rev. Lett. {\bf 97}, 070402 (2006).

\bibitem{Footnote} This result is obtained by assuming the
universal form $P=P_N f(\eta)$, with $n_i=\partial
P/\partial\mu_i$, and using the continuity of pressure and
chemical potentials at the transition.

\bibitem{Zwirlein06c} M.W. Zwierlein and W. Ketterle, e-print
cond-mat/0603489.

\bibitem{Lobo06} We note that the upper bound found here is very
close to the result of recent Monte Carlo simulations presented in
C. Lobo  C. Lobo, A. Recati, S. Giorgini, and S. Stringari,
e-print cond-mat/0607730.

\bibitem{Combescot06} R. Combescot, private communication, to be
published.



\end{thebibliography}
\end{document}